\documentclass[12pt]{article}
\usepackage[usenames,dvips]{pstcol}
\usepackage{color}
\usepackage{amssymb}
\usepackage{epsfig}
\usepackage{footnote}
\usepackage{longtable}
\usepackage{subfigure}
\usepackage{verbatim}
\usepackage{lineno}
\usepackage{setspace}
\usepackage{epstopdf}
\def\Al2O3{Al$_2$O$_3$}

\def\18F{^{18}F}

\newrgbcolor{maroon}{.45 0.1 0.1}
\begin{document}
\pagestyle{plain}

\begin{flushright}
Version v6d (accepted manuscript)\\
\today
\end{flushright}
%==================================================================================
%==================================================================================

%

% Title
%
%\section{Title}
\begin{center}
{\Large Surface Direct Conversion of 511 keV Gamma Rays in Large-Area  Laminated
Multichannel-Plate Electron Multipliers}\\

\vspace*{0.25in}

Kepler Domurat-Sousa, Cameron Poe, Henry J. Frisch\\
{\it Enrico Fermi Institute, University of Chicago}\\
\vspace*{0.05in}

Bernhard W. Adams\\
{\it Quantum Optics Applied Research}\\
\vspace*{0.05in}

Camden Ertley\\
{\it SouthWest Research Institute}\\
\vspace*{0.05in}

 Neal Sullivan\\
{\it Angstrom Research, Inc}\\
\vspace*{0.05in}

\vskip 0.1in
 {\it  Published in Nuclear Instruments and Methods, Section A}
\end{center}

\vskip 0.15in

\begin{abstract}
We have used the TOPAS simulation framework to model the direct
conversion of 511 keV gamma rays to electrons in a micro-channel plate
(MCP) constructed from thin laminae of a heavy-metal-loaded dielectric
such as lead-glass, patterned with micro-channels (LMCP). The laminae serve as
the  converter of the gamma ray to a primary electron within a depth
from a channel-forming surface such that the electron penetrates the
channel surface (`surface direct conversion'). The channels are coated
with a secondary-emitting material to produce electron multiplication
in the channels. The laminae are stacked on edge with the channels
running from the top of the resulting `slab' to the bottom; after
assembly the slab is metalized top and bottom to form the finished
LMCP.

The shape of the perimeter of a lamina determines the dimensions
of the slab at the lamina location in the slab, allowing non-uniform
cross-sections in slab thickness, width, and length.  The slab also
can be non-planar, allowing curved surfaces in both lateral dimensions.
The laminar construction allows incorporating structural elements in
the LMCP for modular assembly in large-area arrays.

The channels can be patterned on the laminae surfaces with internal shapes
and structure, texture, and coatings optimized for specific
applications and performance.  The channels can be non-uniform across
the LMCP and need not be parallel in either transverse direction.

Surface direct conversion of the gamma ray to an electron eliminates
the common two-step  conversion of the gamma ray into an optical photon
in a scintillator followed by the conversion of the photon into an
electron in a photodetector. The simulations predict an efficiency for
conversion of 511 keV gamma rays of $\gtrapprox$ 30\% for a 2.54
cm-thick lead-glass LMCP. The elimination of the photocathode allows
assembly at atmospheric pressure.

\end{abstract}

\setcounter{secnumdepth}{4} % 3 is subsubsection 4 is paragraph
\setcounter{tocdepth}{4}

%\newpage
%\tableofcontents
\newpage

\section{Introduction}

\begin{figure}[!bh]
  \centering
  \includegraphics[angle=0,width=0.80\textwidth]{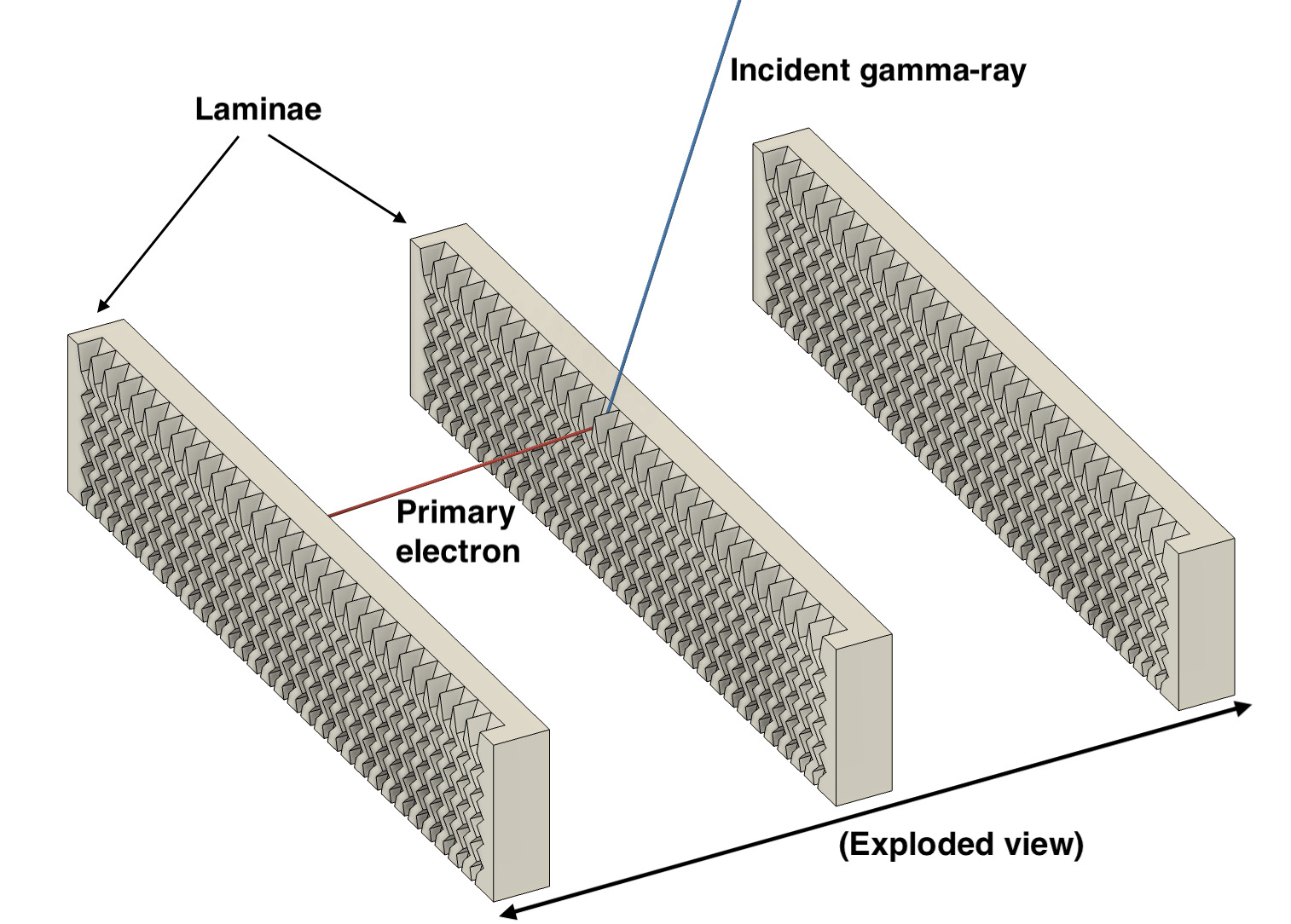}
  \caption{A 3D exploded depiction of three laminae of an LMCP. A 511 keV gamma ray (blue line) produces a primary electron (red line) at a surface in one of the channels. This specific LMCP has micro-patterned channels  with entrance funnels. Note: The image is not to scale, and the particle kinematics have not been modeled here; the depicted primary electron direction is arbitrary.}
   \label{fig:3d_laminae_gamma_incoming}
\end{figure}

The ability to detect low-energy gamma rays with sub-mm space
resolution in three dimensions and time resolutions less than
$\approx$100 ps would open opportunities for simpler, more-capable
detectors in medical imaging, particle physics, nuclear physics,
astrophysics, and other fields. Low-energy gamma rays are now typically
detected by the two-step conversion of the gamma ray to optical photons
in a scintillating medium followed by photon conversion to electrical pulses
in a photodetector~\cite{Vandenberghe_Moskal_Karp_review_2020,Vaquero_Kinehan_review_2015,Phelps_Cherry_Dahlbom_book_2006,
Vandenberghe_Moskal_Karp_2020_Whole_Body_PET_2020,Cherry_Explorer_scattering_2019}.

To explore alternatives at the level of the underlying physics
processes, we have modeled the direct conversion of 511 keV gamma rays
to electrons in a laminar micro-channel plate (LMCP) constructed from
thin heavy-metal-loaded dielectric laminae such as lead-glass.  The
laminae serve as the  converter of the gamma ray to a primary electron
within a depth from a channel-forming surface such that the electron
penetrates the interior channel surface (`surface direct conversion',
or `SDC'). The channels are coated with resistive and
secondary-emitting (SEY) material~\cite{Slade_SEY_NIM} to produce electron multiplication in the
channel much as in a conventional MCP~\cite{Wiza}. A 3D visualization of an LMCP is depicted in Figure~\ref{fig:3d_laminae_gamma_incoming}.

The laminar design with stacked patterned laminae allows wide
flexibility in the 3D shape of the LMCP, and the shapes and
functionalization of the channels. Additionally, the absence of a
photocathode allows modular pre-assembly of large-area arrays of LMCPs
at atmospheric pressure. Minor modifications to the perimeters of the
laminae can provide precision alignment and internal support against
atmospheric pressure in large-area arrays in a single hermetic vessel.

This paper focuses on the surface direct conversion of 511 keV gamma
rays. However, the laminar construction can be applied to a wide
variety of MCP applications, from conventional uniform rectangular MCPs
to non-rectangular and non-planar assemblies, with  locally optimized
non-uniform channel shape, size, direction, coatings, and
metalizations. Substrates can be as in conventional MCPs, or
appropriately optimized for charged particles, neutrons, or
high-energy particle showers in a given energy range.

The organization of the paper is as follows. Section~\ref{LMCP_construction} is a
description of the construction of the laminar MCP.
Section~\ref{lamina} gives an overview of the lamina, including the
substrate material, shape, orientation, and use as a structural
element. Section~\ref{microchannels} presents the patterned
microchannels on the lamina surfaces, including methods of patterning,
channel shapes, and coating methods. Section~\ref{slab_assembly}
describes the assembly of the laminae into a `slab', which when
metalized top-and-bottom becomes the LMCP.  Possible slab geometries
include planar and non-planar slabs, slabs non-uniform in thickness,
width, or length, and slabs with non-parallel channels.  Section~\ref{simulation}
presents the results of  a TOPAS Geant4-based simulation of surface
direct production of electrons from 511-keV gamma rays in tungsten and
a heavy-metal-loaded glass.

\section{Construction of the Laminar MCP (LMCP)}
\label{LMCP_construction}

The laminated microchannel plate (LMCP) is formed by assembling thin
laminae, rather than the conventional method of assembling and drawing
fibers~\cite{Wiza}. An LMCP channel plate is assembled from laminae
with one or both sides patterned with channels that extend from one
edge of the lamina (corresponding to the top surface of the LMCP) to
the other edge (the LMCP bottom). The channels can be patterned in both
transverse and longitudinal directions. Options include entrance
funnels, patterns of strike surfaces, and an exit `choke,' such as seen in Figure~\ref{fig:3d_laminae_gamma_incoming}.

If the substrate is not intrinsically resistive, the laminae can first
be functionalized on both sides with a resistive layer. A
secondary-emitting layer is then applied. Subsequent metalization
before assembly can provide customized dynode structures in the
channels.

After patterning and functionalization, the laminae are stacked, much
like books on a book-shelf, and formed into the plate, here called a
`slab' to distinguish it from a fully-finished LMCP.  The slab can vary
in any of the three dimensions and be non-planar; the channels need
only connect the top and the bottom, but can vary in density, size,
channel-defining surfaces, functionalization, and direction within a
single slab.

To form the finished LMCP, the top and bottom surfaces of the slab are
appropriately metalized to supply voltage and current for electron
amplification in the channels.
%
%Figure 6 Slab made from laminae of heavy-metal glass such as %Pb-glass (Pb-G), ceramic, or other dielectric. The thicknesss of the lamina is $\tau$; the width of %the lamina is equal to the %thickness of %the slab $T$.  The length of the lamina $L$ is %equal to the width/length of the slab.
%
\begin{figure}[!th]
\centering\includegraphics[angle=0,width=0.80\textwidth]{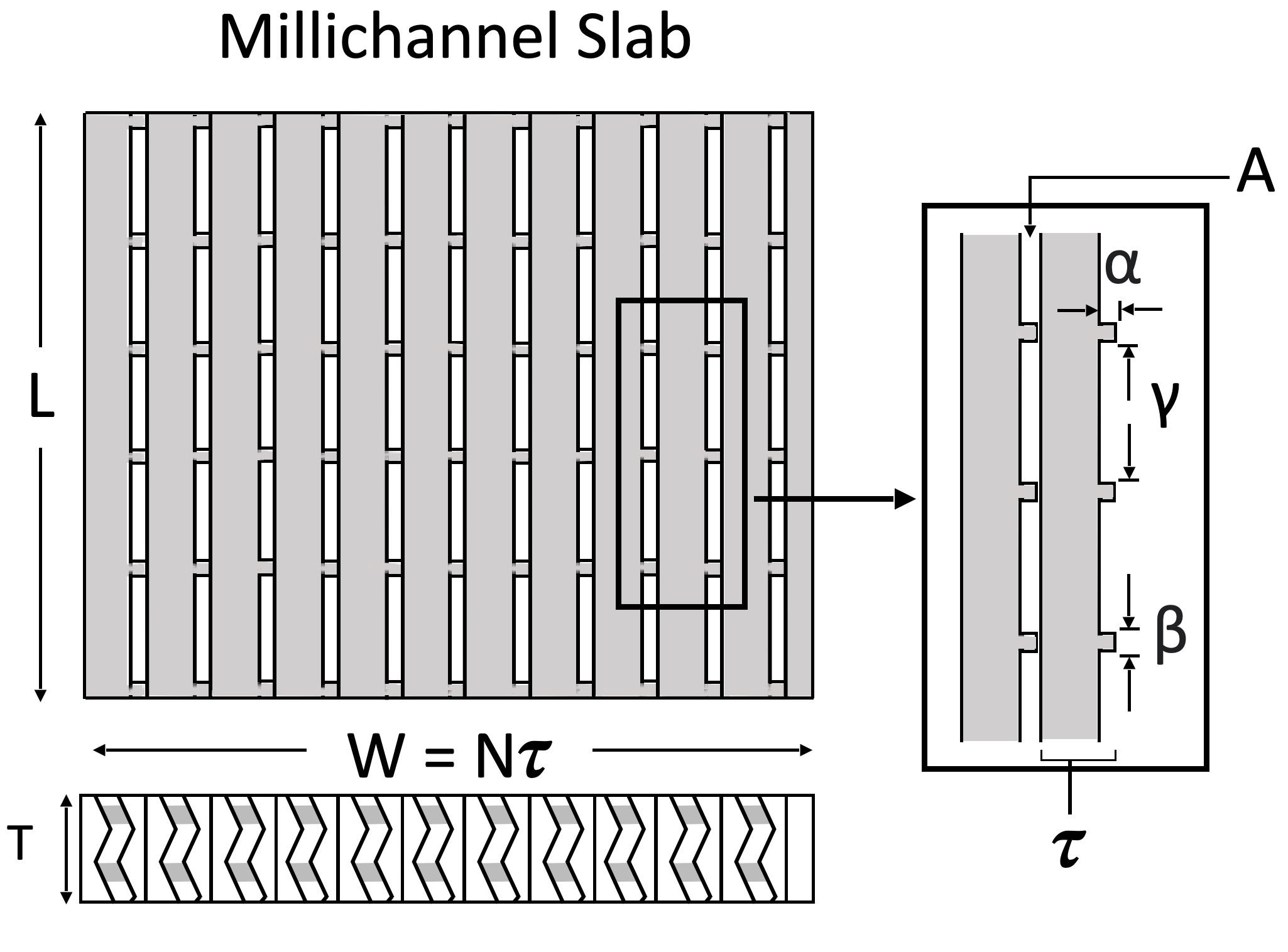}
\caption{A `slab' intended for gamma ray detection made from  a
heavy-metal dielectric such as lead-glass. The slab is formed from
parallel laminae of thickness $\tau$, width $T$, and length $L$, with
ridges of height $\alpha$.  Laminae are stacked on
edge such that the ridges form channels between the laminae. 
The bulk laminae, which represent the largest fraction
of the area of the slab, serve to convert the incoming gamma ray to an
electron.} \label{fig:slab_w_ridges}
\end{figure}

%Figure 6
 Figure~\ref{fig:slab_w_ridges} shows a
functional diagram of a lead-glass slab with dimensions
appropriate for gamma ray conversion to electrons. Both sides of the
laminae are functionalized with resistive and SEY layers. The values of 
the channel-defining dimensions $\tau$ and $\beta$
should be comparable to the range of the primary electron from the
gamma ray, e.g. 150 microns in lead-glass.

The slab outer dimensions and the shape are free to be determined
by operational considerations such as the function, desired packing
fraction of modules in arrays, mechanical assembly, or electrical
connections. In addition to thicker slabs, multiple LMCPs, with or
without an associated anode, can be stacked in series to increase the
detection efficiency.

\section{The Lamina Substrate}
\label{lamina}

In addition to supporting the patterned channels, the lamina substrate
performs three functions: (1) providing the target material for the
production and escape of a primary electron from a gamma ray
conversion; (2) determining the cross-sectional shape of the slab at
the location of the lamina in the slab assembly; and (3) providing
mechanical support and alignment mechanisms for the internal stack.

The individual shapes of the laminae determine the
local cross-section of the slab. For a planar LMCP, as is
typical in current commercial MCPs, the lamina are rectangular and
identical. For a curved LMCP, one or more of the lamina edges will
be curved. The dimensions of the laminae do not need to be the same
throughout the stack; the lateral dimensions and the width of the LMCP
can be varied throughout the slab. Following the analogy of the laminae
stack corresponding to books on a bookshelf, the laminae may be tilted
to one side, as so often happens when a bookshelf is not full.

The ability to independently shape each lamina allows incorporating the
laminae as structural elements in the mechanical package. The edges of
selected laminae can be shaped with protruding `tabs' such that when
the laminae are incorporated on edge into the slab the tabs form
spacers. The spacers separate components inside the assembled detector,
for example between one LMCP and another, or between the top of the top
LMCP and a window or top plate of the hermetic package. The tabs on
successive slabs can form a column between the top and bottom plates to
support atmospheric pressure over large areas. Other features on the
lamina edges  can provide mechanical mounting and/or aligning
mechanisms, and can also be conducting or resistive to be part of the
LMCP internal electrical circuit.

Special laminae made of a non-conducting material chosen for strength,
such as alumina, can be included in the slab at intervals. These
`strong-backs' can be structural elements in the internal mechanical
assembly to support atmospheric pressure, for example.

\section{Micro-Channels on the Lamina Surfaces}
\label{microchannels}

The laminae can be laid out flat for patterning, functionalization,
and metalization of the channels prior to assembly. The laminae can be
micro-patterned by many conventional methods, including
additive/subtractive processes such as 3D printing. The cross-section,
surface texture, and dimensions can change along the channel.

Access to the full channel surface before assembly allows customizing
the interior of the channels for parameters such as gain, time
resolution, pulse uniformity, and rate capability. There are  many
options for secondary coatings, including an option of successive
coatings with selective masking. Metalization also be patterned  along
the channel to produce, for example, custom-spaced dynode structures
and end-spoiling.

\begin{figure}[!th]
\centering
\includegraphics[angle=0,width=0.80\textwidth]{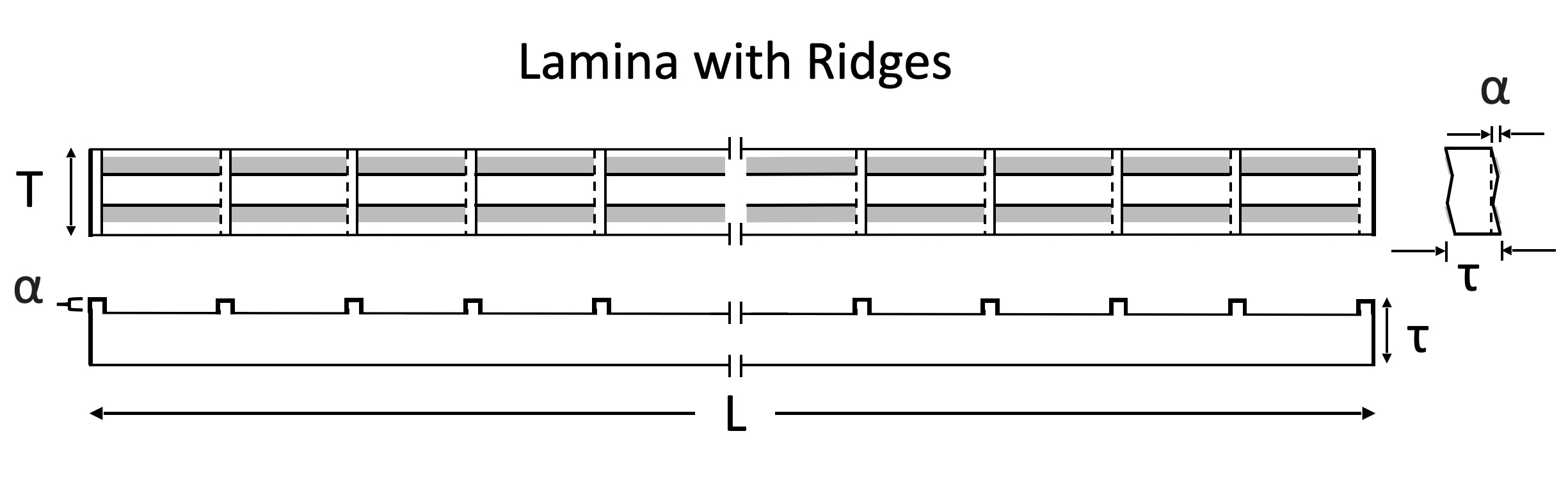}
\caption{An illustrative lamina with a simple pattern of linear
ridges of height $\alpha$, width $\beta$, to
provide the channels. To form a slab, the laminae are stacked on edge
next to each other, much as microscope slides stacked on edge in a box
form a rectangular slab. In the slab the ridges form gaps between the
lamina, creating rectangular channels of dimension $\alpha$
perpendicular to the laminate and $\beta$ along the laminate.
 } \label{fig:lamina_w_ridges}
\end{figure}

%Figure 7
Figure~\ref{fig:lamina_w_ridges} shows a simple regular rectilinear
pattern of channels printed on a lamina that forms  channels from the
top of the slab to the bottom. The channels can be patterned with a
bias angle such that two superposed slabs form a chevron.

%
%Figure 7 A Lamina w straight ridges of a Laminated Converter-Amplifier Slab
%

%
%Figure 8 Segment of Lamina w funnels, dynodes, and end-spoiling of a Laminated Converter-Amplifier Slab
%

\begin{figure}[!th]
\centering
\includegraphics[angle=0,width=0.80\textwidth]{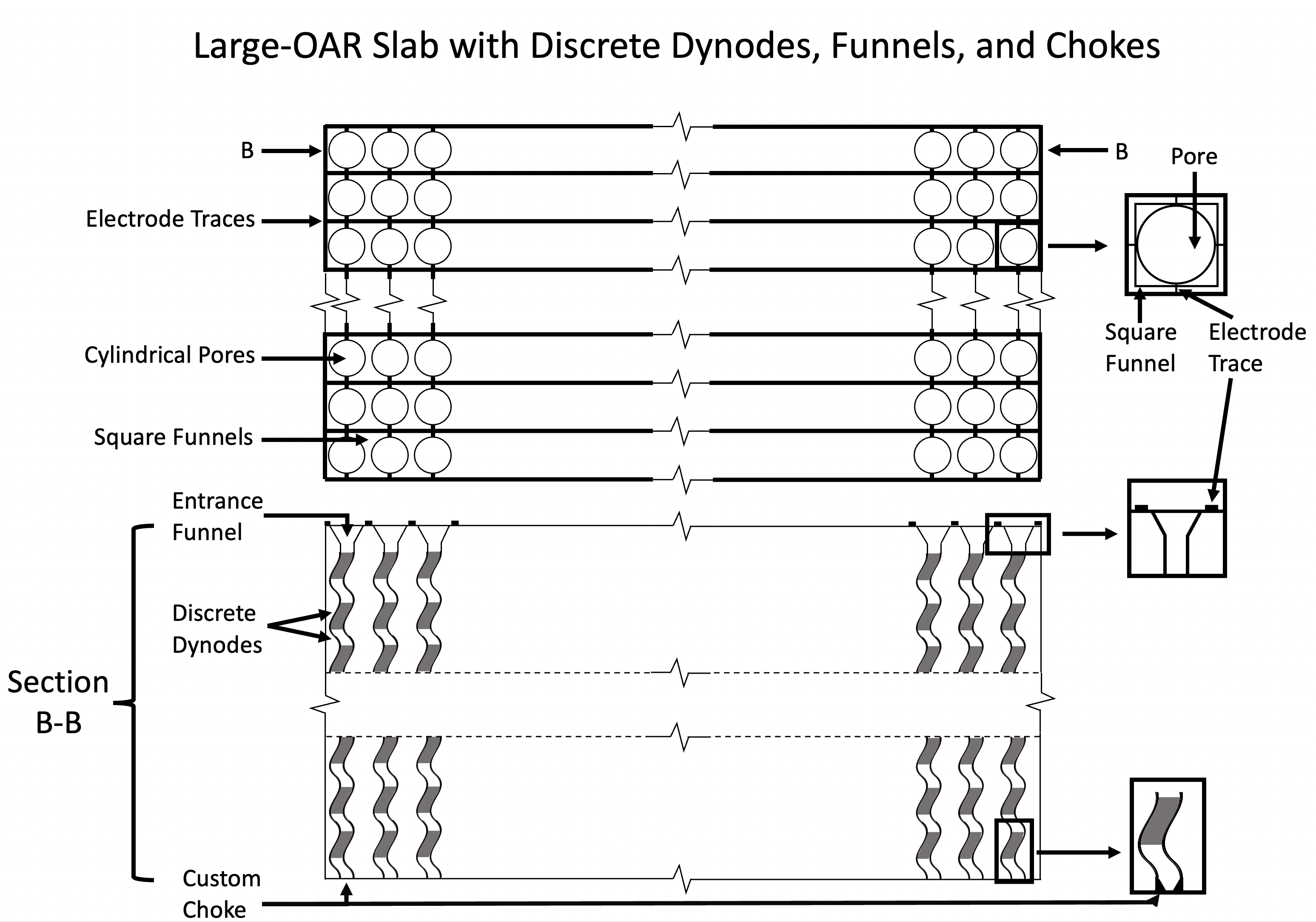}
\caption{A slab composed of laminae with an example pattern based on
cylindrical channels with a square entrance funnel, discrete metalized
dynodes, a mild exit `choke', and exit end-spoiling. The entrance
funnels cover the entrance surface except for thin metalized traces
that supply the voltage and current to each channel, resulting in a
large effective open-area ratio.} \label{fig:slab_w_funnels}
\end{figure}

%Figure 4
Figure~\ref{fig:slab_w_funnels} shows a slab composed of laminae with
an example pattern based on cylindrical channels with an entrance
funnel, discrete metalized dynodes, a mild exit `choke', and exit
end-spoiling. The square entrance funnels cover the
entrance surface except for the interwoven metalized traces that supply
the voltage and current to each channel, resulting in a large open-area
ratio.

\section{Assembling the Laminae}
\label{slab_assembly}

After functionalization, the laminae are bonded together into a solid
slab. The laminar construction allows forming non-uniform and
non-planar slabs. The laminae can be non-rectangular to form a desired
shape transverse to the stacking direction. The laminar construction
also allows making slabs of non-uniform thickness by changing the
dimensions of the laminae during the assembly of the stack. Precision
locating holes or external reference grooves can be used to enforce
precision mating of the two channel patterns on neighboring laminae. In
addition to this flexibility in shape and thickness of the slab,
laminar construction enables non-parallel channels and/or channels of
varying dimensions, as may be desirable from local rate considerations
or for directing the exit shower.

Once the laminae have been bonded together into a slab, the top and
bottom surface of the slab can be metalized to provide the electrodes
to supply voltage across the channels and current through the surface
of the channels. The thin traces between the funnels in the plan view
of Figure~\ref{fig:slab_w_funnels} are an example top electrode.

\section{Simulation using the TOPAS Geant4-based Package}
\label{simulation}

In order to test the viability of such laminar MCP geometries, we have modified~\cite{TOPAS_methods_paper} the TOPAS
package~\cite{TOPAS} to study their efficiencies. Special attention has been paid to the efficiency of direct
conversion of 511-keV gamma rays via surface direct conversion in
laminae containing high-atomic number nuclei such as lead (Pb) or
tungsten (W) for Time-of-Flight Positron-Emission Tomography (TOF-PET).

\subsection{Direct Gamma Conversion to Electrons: Compton
Scattering and the Photoelectric Effect}

%Figure 1 Electron Yield by process vs angle for 50 microns of W and 150 microns of Pb Glass
%
\begin{figure}[!bh]
\centering
\includegraphics[angle=0,width=0.80\textwidth]{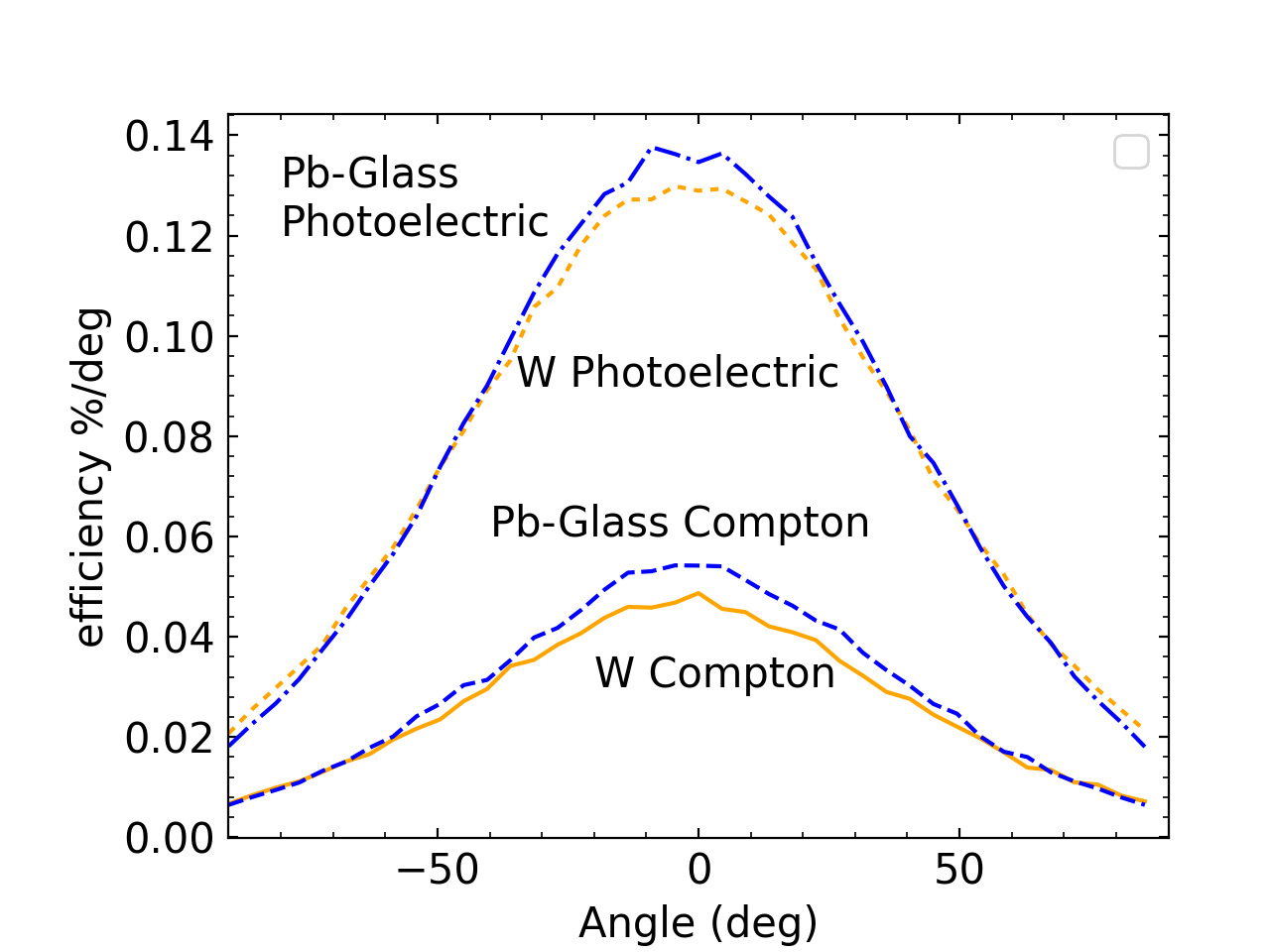}
\caption{The predicted electron angular distribution from the exit side
of a thin plate from an incident
 511 keV gamma ray at normal incidence. The plate thickness for
 lead-glass ($\lambda_R=1.265$ cm)  is 150 microns (0.006")
 and for  W ($\lambda_R=0.350$ cm)  is 50 microns (0.002").}
 \label{fig:yield_vs_angle_compton_PE_W_PbG_511}
\end{figure}

%
%Figure 2 Electron Yield by process vs energy for 50 microns of W and 150 microns of Pb Glass
%
\begin{figure}[!th]
  \centering
  \includegraphics[angle=0,width=0.80\textwidth]{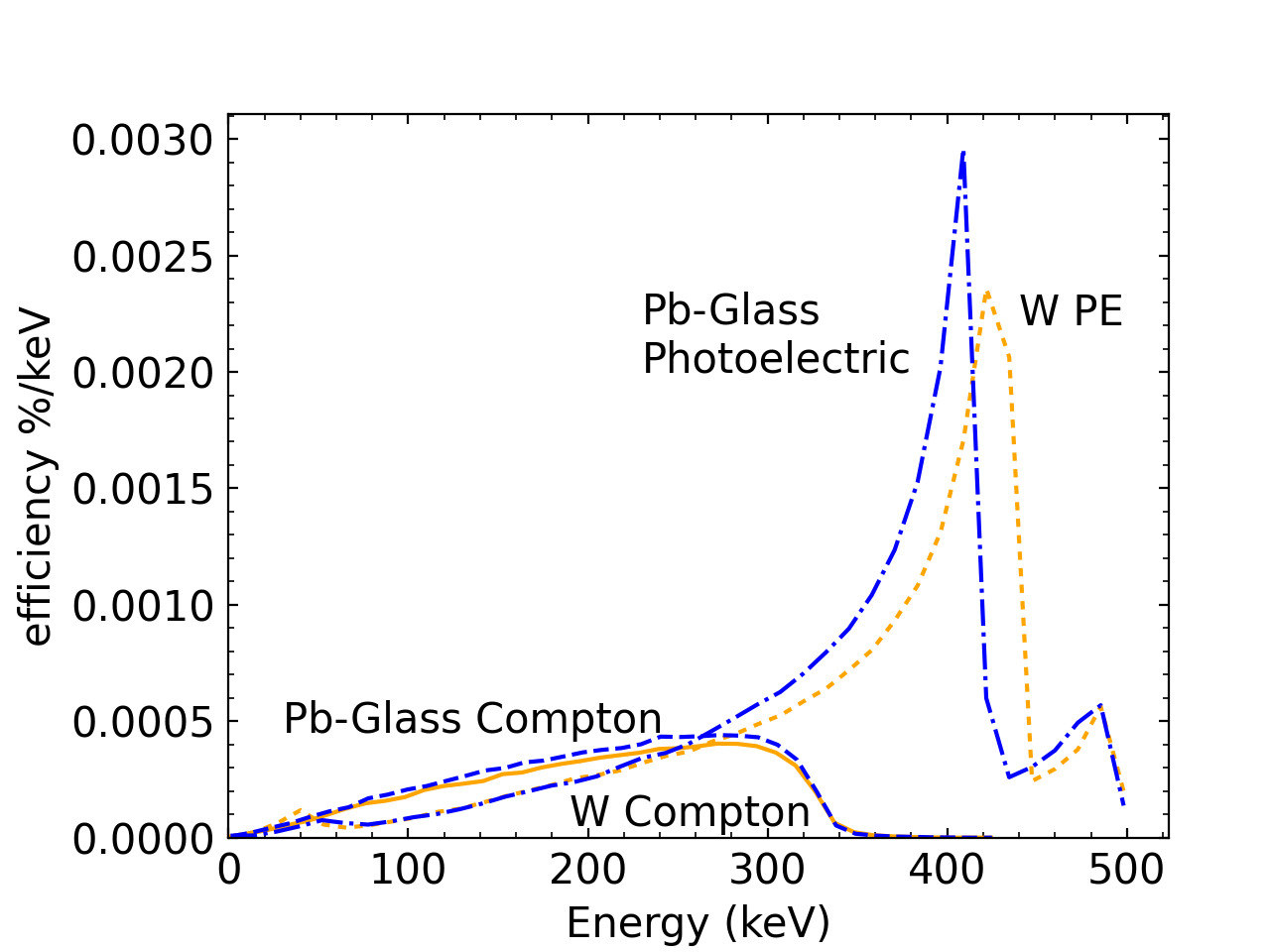}
  \caption{The predicted electron energy spectrum from the exit side of a
  thin plate from an normally incident
   511 keV gamma ray.  The plate thickness for lead-glass ($\lambda_R=1.265$ cm)  is  150 microns (0.006")
   and for  W ($\lambda_R=0.350$ cm)  is 50 microns (0.002").}
   \label{fig:yield_vs_energy_compton_PE_W_PbG_511}
  \end{figure}

The three fundamental processes for conversion of a gamma ray to
charged particles are the Photoelectric Effect, Compton Scattering, and,
at gamma ray energies above 1 MeV, Pair Production. The capability to
convert a sub-MeV gamma ray directly into an electron, which is then multiplied
in the LMCP, has the disadvantage that the efficiency is inherently low
per unit length of material.

The efficiency for a gamma ray to create an electron cascade in a
channel for a given material is determined by two factors. First, the
cross sections and energy distributions for producing  a `primary'
electron depend on the substrate. Second, the range of sub-MeV
electrons in heavy materials is short; only conversions close to the
functionalized channel interior surface reach it before ranging out.
These production and escape factors vary with material: tungsten (W),
for example,  has a high conversion cross-section, but a short escape
depth. Lead-glass, for example, has a lower conversion cross-section
but a longer escape depth. The values of the radiation lengths ($\lambda_R$) assigned by TOPAS in the simulation were 1.265 cm for lead-glass (`G4\_GLASS\_LEAD') and 0.350 cm for W (`G4\_W'). 

Figure~\ref{fig:yield_vs_angle_compton_PE_W_PbG_511} shows the
predicted electron angular distributions from the exit side of a thin
plate from a 511 keV gamma ray at normal incidence.  The plate
thickness for lead-glass is 150 microns (0.006") and for  W is 50 microns
(0.002").

%\clearpage

%Figure 2 description
The predicted electron energy spectrum from the exit side of a thin
plate from a 511 keV gamma ray at normal incidence is shown in
Figure~\ref{fig:yield_vs_energy_compton_PE_W_PbG_511}.  The respective plate
thicknesses are  150 microns (0.006") and 50 microns (0.002")  for lead-glass and W.
The structures below 500 keV are due to the atomic K and L shells.

%\clearpage
\subsection{Surface Direct Production of a Primary Electron}
%
%Figure 3: Electron Yield vs thickness for W and lead_glass at 511 keV thin plate
%
\begin{figure}[!th]
\centering
\includegraphics[angle=0,width=0.80\textwidth]{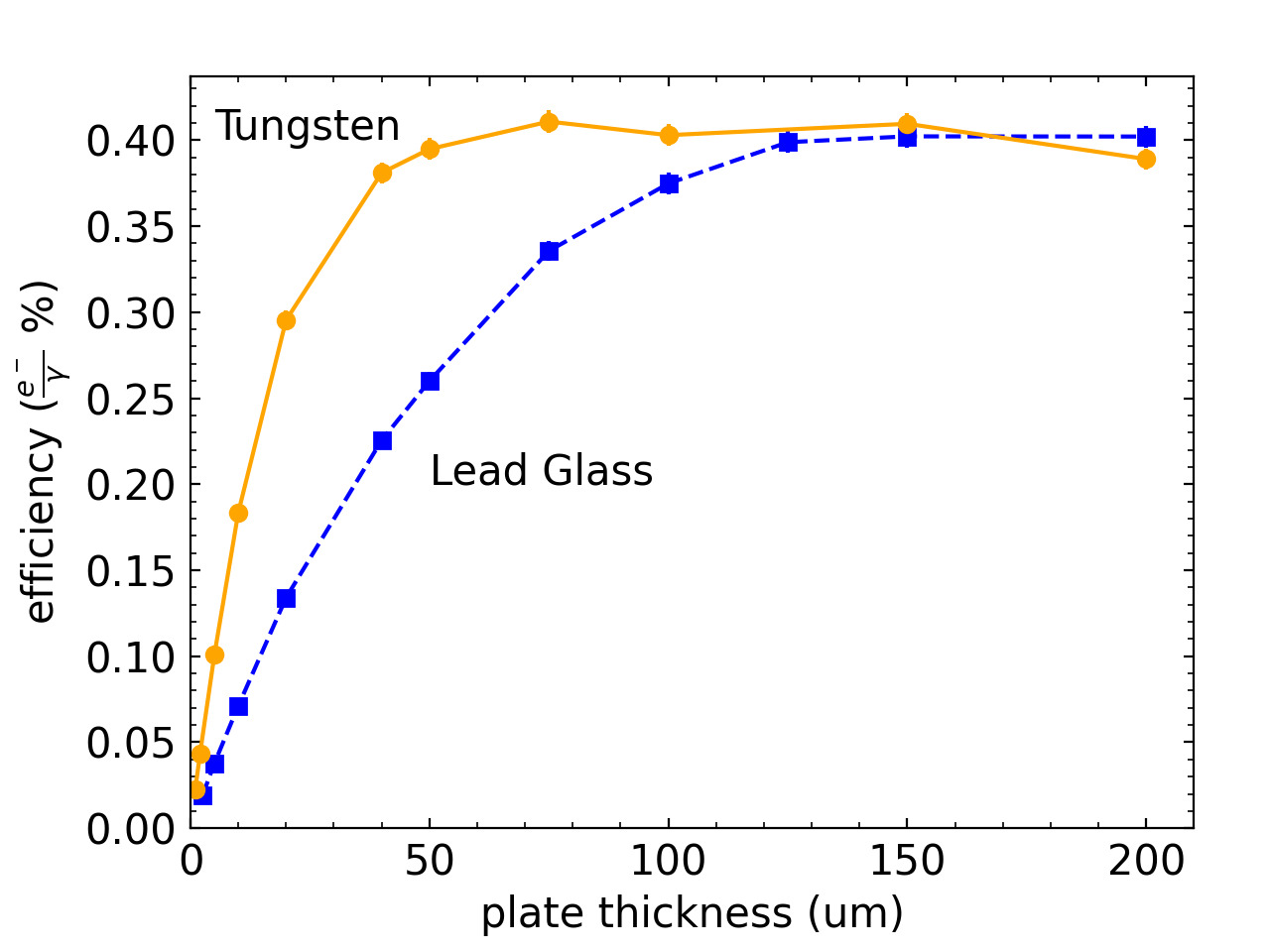}
\caption{The predicted electron yield per 511 keV gamma ray traversing
a thin sheet of tungsten (W) (solid) and lead-glass (dashed) at normal
incidence versus plate thickness in microns. The yield is for only the
exit side of the plate. The lower density of the glass is compensated
by the larger depth from which electrons reach the surface. We take as
nominal values for the thickness at which the curve has reached its
plateau as 150 microns for lead-glass and 50 microns for W.}
\label{fig:yield_vs_thickness_W_PbG_511}
\end{figure}

%Figure 3 description:
The predicted electron yield per 511 KeV gamma ray traversing a thin
sheet of lead-glass (dashed) and W (solid) at normal incidence versus plate thickness
in microns is shown in Figure~\ref{fig:yield_vs_thickness_W_PbG_511}.
The yield is for the exit side of the plate. The lower density of the
glass is compensated by the larger depth from which electrons reach the
surface. We take as nominal values for the thickness at which the curve
has reached its plateau as 150 microns for lead-glass and 50 microns for
W.

%\clearpage

%\subsubsection{Efficiency Versus Path Length Inside a Thin Lamina}
%%
%Figure 4: Yield vs path length inside a 2 mil thick W (solid) and 6 mils PbG (dashed) lamina
%
\begin{figure}[!th]
\centering
\includegraphics[angle=0,width=0.80\textwidth]{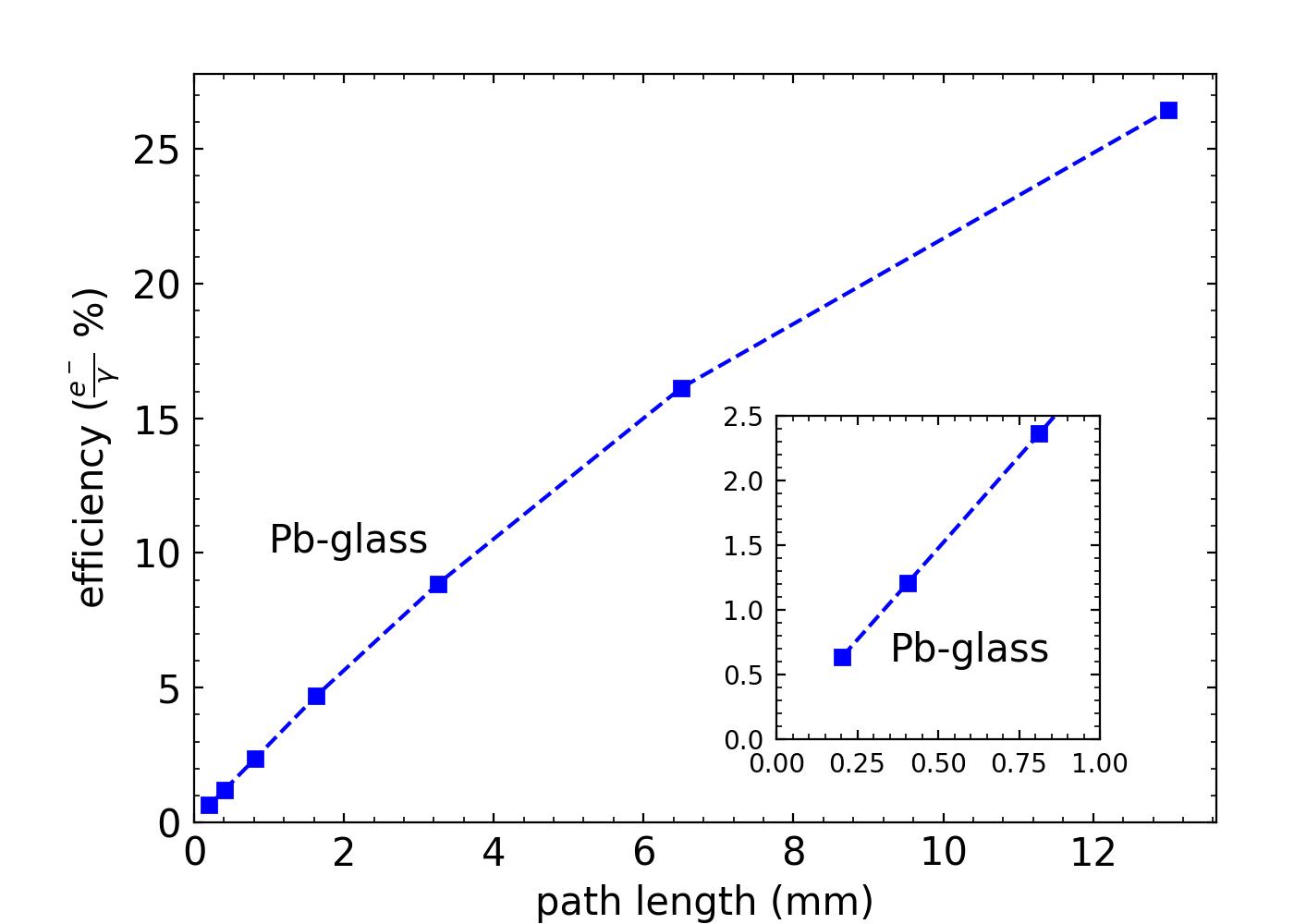}
\caption{The electron yield in percent vs projected path length in a 150 micron
(0.006") lead-glass lamina for a gamma ray propagating inside the lamina substrate.
The yield includes electron emission from both sides of the lamina.
Installed in a slab,  the lamina would be coated with a secondary
emitter such as MgO, \Al2O3, or CVD diamond that the primary electrons
would traverse at the channel surfaces.}
\label{fig:yield_vs_pathlength_6mil_PbG}
\end{figure}

%
%Figure 4 description
Figure~\ref{fig:yield_vs_pathlength_6mil_PbG} shows the efficiency for
the production of an electron that exits a lamina through either
surface versus the projected path length of a 511 keV gamma ray travelling inside
a lead-glass lamina substrate of thickness 150 microns (0.006"). Note
that the efficiency in lead-glass is 25\% at a slab thickness of 1.27 cm
(1/2") and grows approximately linearly, as one would expect.

%\clearpage
%
% Figure 5 Efficiency of 511 keV gamma conversion vs angle from normal in the phi direction (cylinder)
%
\begin{figure}[!th]
\centering
\includegraphics[angle=0,width=0.88\textwidth]{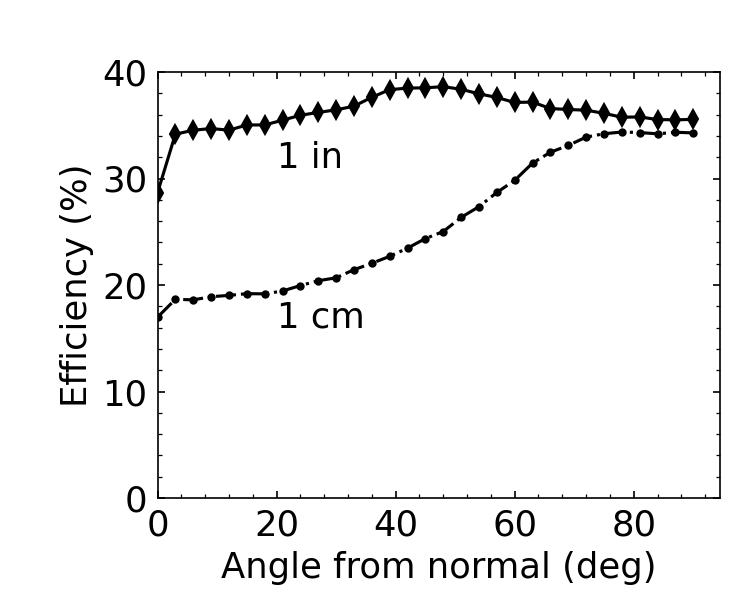}
\caption{The efficiency found in the TOPAS simulation for direct
conversion of a 511 keV gamma ray versus incident angle $\phi$ from the
normal to the LMCP. The efficiency includes the creation
of a primary electron that enters a channel by crossing a
functionalized channel-defining wall. Distributions are shown for MCP
thicknesses of 1 cm and 2.54 cm (1 inch).}
\label{fig:eff_by_ang_phi}
\end{figure}

%
%Figure 5 description;  Electron Yield vs angle for 1-cm and 1-inch 0.002" W, and 0.006" Pb-Glass (PbG) at 511 keV
Figure~\ref{fig:eff_by_ang_phi} shows the efficiency for 1 cm-thick and
2.54 cm-thick LMCPs for conversion of a 511 keV gamma ray to a primary
electron that traverses a channel wall into the channel interior,
crossing at least one layer of secondary-emission material, versus incident gamma ray angle.
Thinner LMCPs will have better time resolution; thicker will have
higher efficiency.

\section{Summary}
\label{summary}
% Surface Direct Conversion

We have studied a laminated microchannel plate (LMCP) formed by assembling thin
laminae, rather than the conventional method of assembling and drawing
fibers. An LMCP channel plate is assembled from laminae
with one or both sides patterned with channels that extend from one
edge of the lamina (corresponding to the top surface of the LMCP) to
the other edge (the LMCP bottom). The channels can be patterned in both
transverse and longitudinal directions. Options include entrance
funnels, patterns of strike surfaces, and an exit `choke.'

We have used the TOPAS simulation framework to determine the parameters
and efficiency for the surface direct conversion of 511 keV gamma rays
to electrons in a micro-channel plate (LMCP) constructed from thin
laminae of lead-glass patterned with micro-channels. Direct conversion
of the gamma ray to an electron eliminates the common two-step  process
of first converting of the gamma ray into an optical photon in a
scintillator, followed by the conversion of the photon into an electron
in a photodetector. The simulations predict an efficiency for
conversion of 511 keV gamma rays of $\gtrapprox$ 30\% for a single 2.54
cm-thick lead-glass LMCP; multiple units can be stacked for higher
efficiencies. The elimination of a photocathode allows assembly at
atmospheric pressure.

The shape of each thin lamina determines the local dimensions of the
LMCP, allowing non-uniform cross-sections in slab thickness, width, and
length. The slab can be non-planar, allowing curved surfaces in both
lateral dimensions.

The channels can be patterned on laminae surfaces with customized
internal shapes, textures and structure. The channel-forming surfaces
can be functionalized with resistive, secondary-emissive, and
conducting coatings with patterns and thicknesses selected for specific
applications. The channels need not be parallel nor uniform across the
LMCP, for example in concave/convex LMCPs for focusing, or LMCPs in
rapidly varying rate environments such as close to an accelerator beam.

The laminar construction allows incorporating structural elements
directly in the LMCP. Tabs can be added to the lamina perimeter to
provide precision alignment and internal support against atmospheric
pressure in large-area multi-module vacuum vessels such as planes of
pre-shower samplers in particle physics, or in large area radially-thin
cylindrical geometries in TOF-PET.

%Figure 10 HGM Stackup w 2 Slabs
%
\begin{figure}[!th]
  \centering
  \includegraphics[angle=0,width=0.80\textwidth]{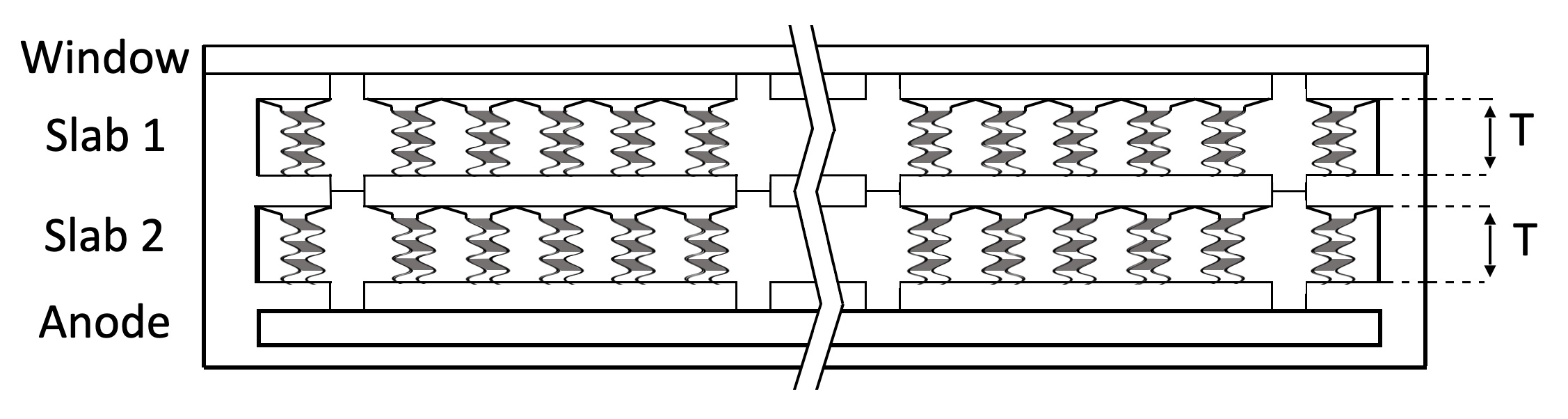}
  \caption{An example detector assembly comprising two LMCPs formed with
  laminae that form electron-multiplier channels with entrance funnels.
  The effective open-area ratio (OAR) is large due to the funnels, while
  the smaller channels maximize the area presented to gamma rays for
  conversion. The laminae are shown with tabs to space the slabs and to
  provide mechanical support in columns between the top window and the
  bottom of the hermetic package.    }
  \label{fig:2_slab_millichannel_detector}
  \end{figure}

Figure~\ref{fig:2_slab_millichannel_detector} shows an example detector
assembly with two LMCPs in series. We note that as the primary electron
from the gamma ray may cross a channel at any point along the channel
length, the gain in a slab varies event-to-event. This  variable gain
can be ameliorated by positioning the dynodes at the exit end of the
channel, or by adding a conventional non-converting MCP operated in
saturation as the final amplifier in the stack of components.

Lastly, in the assembly of the LMCP into a functional detector, an
anode, either internal~\cite{timing_paper} or
capacitively-coupled~\cite{InsideOut_paper}, with sub-mm spatial
resolution if needed~\cite{patterned_anode_paper}, receives the pulse
from the electron shower, and typically transmits it to fast
digitization electronics~\cite{JF_NIM,PSEC4,Oberla_Clermont_2014}.

\section*{Acknowledgments}
We thank Joseph Perl for the exemplary development of TOPAS and for
remarkable user support. We are indebted to Mary Heintz for essential
computational system development and advice, and to Ian Goldberg and Justin Gurvitch for
crucial graphics contributions.

For financial
support of undergraduate research from University of Chicago College, Physical Sciences Division, and
Enrico Fermi Institute, we thank Steven Balla and Nichole Fazio, Michael Grosse, and Scott Wakely,
respectively.

%\clearpage

%
%%\end{document}

%=========================================================================================================

% THE BIBLIOGRAPHY


\begin{thebibliography}{99}

%


%1
\bibitem{Vandenberghe_Moskal_Karp_review_2020}
S. Vandenberghe, P. Moskal, J. S. Karp;
{\it  State of the art in total body PET}\\
 EJNMMI Phys. 2020 May 25;7(1):35. doi: 10.1186/s40658-020-00290-2.

%2
\bibitem{Vaquero_Kinehan_review_2015} J. J. Vaquero and P. Kinahan; {\it Positron
  Emission Tomography: Current Challenges and Opportunities for
  Technological Advances in Clinical and Preclinical Imaging Systems}
  Annual Review of Biomedical Engineering Volume 17, 385; (2015)

%3
\bibitem{Phelps_Cherry_Dahlbom_book_2006} M. E. Phelps, S. R. Cherry, and M.
Dahlbom;\\
{\it PET: Physics,instrumentation, and scanners}; Springer New York (2006)\\ 
doi.org/10.1007/0-387-34946-4

%4
\bibitem{Vandenberghe_Moskal_Karp_2020_Whole_Body_PET_2020}
S. Vandenberghe, P. Moskal, J.S. Karp;
{\it  State of the art in total body PET}\\
 EJNMMI Phys. 2020 May 25;7(1):35.\\
doi: 10.1186/s40658-020-00290-2. PMID: 32451783; PMCID: PMC7248164.
%note- not a duplicate of their 2015 review

%5
\bibitem{Cherry_Explorer_scattering_2019}
R. D. Badawi, H. Shi, and S. R. Cherry et al.\\
{\it First Human Imaging Studies with the EXPLORER Total-Body PET Scanner};\\
J Nucl Med. 2019 Mar; 60(3): 299-303. doi: 10.2967/jnumed.119.226498
%PMCID: PMC6424228 PMID: 30733314

%6
\bibitem{Slade_SEY_NIM} Z. Insepov, V. Ivanov, S.
    J. Jokela, I. V. Veryovkin and A. V. Zinovev;\\
{\it Comparison of secondary electron emission simulation to
    experiment}; Nucl. Instr. Meth A639, 155 (2011)

%7
\bibitem{Wiza} J.L. Wiza,
Micro-channel Plate Detectors.
Nuclear Instruments and Methods 162, 1979, pp 587-601

%8
\bibitem{TOPAS_methods_paper} K. Domurat-Sousa, C. Poe;
{\it Methods for Simulating Low-Z-Medium-based TOF-PET in TOPAS};
Submitted to Nucl. Instr. and Meth., June, 2023

%9
\bibitem{TOPAS}
 B. Faddegon, J. Ramos-Mendez, J. Schuemann, J. Shin, J. Perl, H.
 Paganetti\\
{\it The TOPAS tool for particle simulation, a Monte Carlo simulation
tool for physics, biology and clinical research}; European Journal of
Medical Physics;  Volume 72, P114-121, April (2020);
DOI:https://doi.org/10.1016/j.ejmp.2020.03.019

%9
\bibitem{timing_paper}B.W. Adams, A. Elagin, H. Frisch, R. Obaid,
E. Oberla, A. Vostrikov, R. Wagner, J. Wang, M. Wetstein; {\it  Timing
Characteristics of Large Area Picosecond Photodetectors}; Nucl. Inst.
Meth. Phys. Res. A. , Vol. 795, pp 1-11 (Sept. 2015)

%10
\bibitem{InsideOut_paper}
E. Angelico, T. Seiss, B. W, Adams, A. Elagin, H. J. Frisch, E. Spieglan;\\
{\it Capacitively coupled pickup in MCP-based photo-detectors using a
  conductive, metallic anode}; Nucl. Inst. Meth. Phys. Res. A. (Oct. 2016)

%11
\bibitem{patterned_anode_paper} Jinseo Park, Fangjian Wu, Evan Angelico, Henry J. Frisch, Eric Spieglan;
{\it Patterned anodes with sub-millimeter spatial resolution for large-area MCP-based photodetector systems}
Nuclear Inst. and Methods in Physics Research, A 985 (2021) 164702;
22 Sept, 2020

%12
\bibitem{JF_NIM}
 J.-F. Genat, G. Varner, F. Tang, H. Frisch;
 {\it Signal Processing for Pico-second Resolution
Timing Measurements};
 Nucl.Instrum.Meth.A607:387-393,Oct., 2009. arXiv:0810.5590

%13
\bibitem{PSEC4}
E. Oberla, J.-F. Genat, H. Grabas, H. Frisch, K. Nishimura, and G. Varner;\\
 {\it A 15 GSa/s, 1.5 GHz Bandwidth Waveform Digitizing ASIC};\\
 Nucl. Instr. Meth. A735, 21 Jan., 2014, 452

%14
\bibitem{Oberla_Clermont_2014} E. Oberla;
{\it PSEC4 waveform sampler and Large-Area Picosecond Photo-Detectors
readout electronics}: Procedings of the Workshop on Picosecond Photon
Sensors, Clermont-Ferrand, 2014. Available at
http://lappddocs.uchicago.edu/documents/243


\end{thebibliography}
\end{document}